\documentclass[%
 prb,
 amsmath,amssymb,
 reprint,%
superscriptaddress,
longbibliography
]{revtex4-2}

\usepackage[utf8]{inputenc}
\usepackage{graphicx}
\usepackage{amsmath}
\usepackage{float}
\usepackage{amssymb}
\usepackage{placeins}
\usepackage{array}
\usepackage{color}
\usepackage{physics}
\usepackage{makecell}
\usepackage{siunitx}
\usepackage{xstring}
\usepackage{multirow}
\usepackage{mathtools}
\usepackage{soul} 
\usepackage{changes}
\usepackage{dcolumn}
\usepackage{bm}
\usepackage{subcaption} 
\usepackage{makecell}
\usepackage{soul}
\usepackage{comment}
\usepackage{hyperref} 
\hypersetup{
    colorlinks=true,
    linkcolor=red,
    filecolor=cyan,      
    urlcolor=blue,
    citecolor=blue,
    bookmarks=true,
    pdfpagemode=FullScreen
}
\usepackage{tikz}
\usetikzlibrary{shapes.geometric, arrows}

\DeclareUnicodeCharacter{0301}{\'{e}}

\definecolor{myblue}{HTML}{004488}
\definecolor{myorange}{HTML}{DDAA33}
\definecolor{myred}{HTML}{BB5566}

\newcommand{\rn}[1]{%
  \textup{\uppercase\expandafter{\romannumeral#1}}%
}

\def\NVClusterM{\text{C}_\text{33}\text{H}_\text{36}\text{N}^\text{-}}

\def\gs{\prescript{3}{}{A}_2} 
\def\tes{\prescript{3}{}{E}} 
\def\sgs{\prescript{1}{}{E}} 
\def\fses{\prescript{1}{}{A}_1} 
\def\sses{\prescript{1}{}{E'}} 

\newcommand{\vibgs}{{^3\widetilde{A}_2}}
\newcommand{\vibtes}{{^3\widetilde{E}}} 
\newcommand{\vibfses}{{^1\widetilde{A}_1}} 
\newcommand{\vibsgs}{{^1\widetilde{E}}} 
\newcommand{\cisgs}{{^1\bar{E}}} 

\def\gwbse{\text{G}_0\text{W}_0\text{-}\text{BSE}}

\definecolor{kc}{rgb}{0.6,0.2,1}

\begin{document}
\preprint{AIP/123-QED}

\title{Excited-State Dynamics and Optically Detected Magnetic Resonance of Solid-State Spin Defects from First Principles}

\author{Kejun Li}
\affiliation{Department of Physics, University of California, Santa Cruz, California, 95064, USA}
\author{Vsevolod D. Dergachev}
\affiliation{Department of Chemistry, University of Nevada, Reno, Nevada, 89557, USA}
\author{Ilya D. Dergachev}
\affiliation{Department of Chemistry, University of Nevada, Reno, Nevada, 89557, USA}
\author{Shimin Zhang}
\affiliation{Department of Materials Science and Engineering, University of Wisconsin-Madison, 53706, USA}
\author{Sergey A. Varganov}
\affiliation{Department of Chemistry, University of Nevada, Reno, Nevada, 89557, USA}
\author{Yuan Ping}
\email{yping3@wisc.edu}
\affiliation{Department of Materials Science and Engineering, University of Wisconsin-Madison, 53706, USA}
\affiliation{Department of Physics, University of California, Santa Cruz, California, 95064, USA}


\begin{abstract}
    Optically detected magnetic resonance (ODMR) is an efficient and reliable method that enables initialization and readout of spin states through spin-photon interface. In general, high quantum efficiency and large spin-dependent photoluminescence contrast are desirable for reliable quantum information readout. 
    However, reliable prediction of the ODMR contrast from first-principles requires accurate description of complex spin polarization mechanisms of spin defects. These mechanisms often include multiple radiative and nonradiative processes in particular intersystem crossing (ISC)  among multiple excited electronic states. 
    In this work we present our implementation of the first-principles ODMR contrast, by solving kinetic master equation with calculated rates from \textit{ab initio} electronic structure methods 
    then benchmark the implementation on the case of the negatively-charged nitrogen vacancy center in diamond. We show the importance of correct description of multi-reference electronic states and pseudo Jahn-Teller effect for quantitatively, even qualitatively correct prediction of spin-orbit coupling (SOC) and the rate of ISC.
    We present the complete calculation of SOC for different ISC processes that align with both group theory and experimental observations. Moreover, we provide a comprehensive picture of excitation and relaxation dynamics, including previously unexplored internal conversion processes.
    We show good agreement between our first-principles calculations and the experimental ODMR contrast under magnetic field. We then demonstrate reliable predictions of magnetic field direction, pump power, and microwave frequency dependency, as important parameters for ODMR experiments. 
    Our work clarifies the important excited-state relaxation mechanisms determining ODMR contrast and provides a predictive computational platform for spin polarization and optical readout of solid-state quantum defects from first principles.
\end{abstract}

\maketitle


\section{Introduction}
Point defects in solids as spin qubits offer multiple avenues to quantum technologies, in particular quantum sensing and quantum networking~\cite{mamin2013nanoscale}. The promising candidates exhibit long quantum coherence times~\cite{ryan2010robust,dolde2014high,gottscholl2021room}, essential for performing multi-step quantum operations with high fidelity. They are relatively scalable and easy to integrate into existing technologies due to their solid-state nature~\cite{bernien2013heralded}. Furthermore, they can be optically initialized and reliably read out in a wide temperature range~\cite{gruber1997scanning,toyli2012measurement,scheidegger2022scanning}, through the spin-photon entanglement~\cite{togan2010quantum}. While discovering and exploring new spin defects beyond the extensively-studied negatively-charged nitrogen-vacancy (NV) center in diamond~\cite{doherty2013nitrogen} has been a key interest in recent years, reliable theoretical predictions of optical readout properties remain challenging, which hinder rapid progress.

The first outstanding issue is the missing link between experimental observable and first-principles simulations for the spin polarization processes. The experimental tool to study spin polarization is through optically detected magnetic resonance (ODMR)~\cite{tetienne2012magnetic,PhysRevB.108.085203,PhysRevLett.131.086903,baber2021excited}, by recording photoluminescence contrast with and without microwave radiation as a function of magnetic field. This requires to solve the kinetic equations for excited-state populations of different spin sublevels. The excited-state population is determined by kinetic processes in the polarization cycle, including radiative and nonradiative recombination between different spin states.  Previously, only model simulations with experimental rates and energy levels to describe ODMR have been reported~\cite{tetienne2012magnetic,PhysRevB.108.085203,PhysRevLett.131.086903,baber2021excited}. A first-principles formalism and computational tool is yet to be developed in order to interpret experiments or predict ODMR of new spin defect systems.

To perform first-principles ODMR, predictions of optical excitation energies, zero-field splitting, radiative and nonradiative recombination rates are required. Various first-principles electronic structure methods have been proposed to calculate excitation energies of spin defects. 
The key consideration is to accurately capture the electron correlation of defect states. 
For instance, mean-field theory such as constrained DFT (CDFT)~\cite{thiering2018theory} could provide reasonable structural and ground state information, but GW and solving the Bethe-Salpeter Equation (GW-BSE) can provide more accurate quasiparticle energies and optical properties including electron-hole interactions~\cite{smart2021intersystem,wu2017first}. On the other hand, if the defect states have strong multi-reference nature, such as open-shell singlet excited state of the NV center, theories like quantum-defect-embedding theory (QDET) for solids~\cite{ma2010excited,sheng2022green} or multi-reference wavefunction methods~\cite{bhandari2021multiconfigurational} may be more appropriate. Note that such theory still requires substantial development in order to study dynamical and spin-orbit properties.

The first-principles theory for excited-state kinetic properties of spin defects has recent advancement as well. For example, 
the radiative recombination rates have been calculated at finite temperature with inputs from many-body perturbation theory for a better description of excitons~\cite{ma2010excited,wu2019dimensionality}. The nonradiative recombination transitions, in particular intersystem crossing (ISC) that requires spin-flip transition and phonon-assisted internal conversion (IC), can be obtained from phonon perturbation to electronic states~\cite{wu2019carrier,smart2021intersystem,thiering2017ab,thiering2018theory}. However, these calculations have not considered the multi-reference nature of the electronic states and complex Jahn-Teller (JT) effects coupled in the kinetic processes.  

In this work, we have developed the first-principles ODMR theory and computational tool to directly predict ODMR contrast without \textit{prior} input parameters as shown in Fig.~\ref{fig:NV_center_ODMR_workflow}(a). We note that such tool is fully general to all solid-state defect systems, although we first use the NV center as a prototypical example. By combining advanced electronic structure methods and group theory analysis,  we provide a complete set of SOC constants for ISC processes which has not been shown before, and a comprehensive picture of excitation and relaxation dynamics of the NV center. We discuss the significance of 
configuration interaction, pseudo JT and dynamical JT effects to the SOC, ISC and IC. The consideration of these effects facilitates 
the interpretation of the experimentally observed ISC rates that essentially contribute to spin polarization. Particularly, we provide a definitive answer to the longstanding question regarding the axial ISC of $\tes\rightarrow\fses$: why symmetry predicts it to be forbidden in the first order~\cite{manson2006nitrogen,maze2011properties,doherty2011negatively}, experiments show it to be nonzero~\cite{goldman2015phonon,tetienne2012magnetic,robledo2011spin}. Our work demonstrates a tight connection between theoretical predictions and experimental observations for various experimental observables, including excitation energy, radiative recombination rate, SOC, ISC, IC and ODMR. Our theoretical results show qualitative or quantitative agreement with experimental data, depending on properties. We note that, for the quantities related to the open-shell excited states, a proper treatment of possible multireference character is necessary. Additionally, our calculated SOC values complement those difficult to measure experimentally. Importantly, we explain why the transition $\fses \rightarrow \sgs$ is predominantly nonradiative by estimating the IC rates between two singlet states. We predict the ODMR contrast as a function of an external magnetic field fully from first-principles, in good agreement with experimental results. We demonstrate that these simulations are invaluable for predicting spin-polarization mechanism and provide control strategy on high fidelity optical initialization and readout.



\section{Computational Methods}
We carry out first-principles calculations of structural and ground-state properties using the open-source plane-wave code Quantum Espresso~\cite{QE}. The NV center defect is introduced into a simple cubic diamond crystal with $3\times3\times3$ supercell size, containing 216 atoms. Only Gamma point is sampled in the Brillouin zone of the defect supercells for all calculations. 
We use the optimized norm-conserving Vanderbilt (ONCV) pseudopotentials~\cite{ONCV1} with the wavefunction energy cutoff of 70Ry. The lattice constant is optimized by using the exchange-correlation functional with the Perdew-Burke-Ernzerhof (PBE) generalized gradient approximation~\cite{PBE1997}. For geometry optimization and excitation energies, we mainly use the range-separated hybrid function proposed by Heyd, Scuseria, and Ernzerhof (HSE)~\cite{heyd2003hybrid, HSE06}, and part of the results using PBE functional are for comparison with HSE, presented in the supplementary material~\cite{supplementary_material} (SM, see also references~\cite{lu2012multiwfn,palummo2015exciton,tamarat2008spin,pl_alkauskas2014first,razinkovas2021photoionization,rayson2008first,jelezko2004observation,tamarat2006stark,li2022carbon,dean1965intrinsic,mariani2020system} therein). The HSE functional has been shown to better describe the electronic structure of the NV center in diamond compared to the PBE functional~\cite{deak2010accurate,gali2009theory,razinkovas2021vibrational}. We calculate the excitation energies using three different methods, constrained DFT (CDFT), single-shot GW plus the Bethe-Salpeter Equation ($\gwbse$), and complete active space self-consistent field method with second order perturbation theory (CASSCF-CASPT2)~\cite{roos2005casscf}.


The many-body perturbation theory calculations are performed by using the WEST code~\cite{govoni2015large,yu2022gpu,rocca2010ab,PRBDario-2012}, with PBE eigenvalues and wavefunctions as the starting point for quasiparticle energies at GW approximation. With the GW approximation, the electronic structure at this level is corrected by quasiparticle self-energy. The dielectric matrix is solved through a spectral decomposition with the dynamical effect incorporated using the Lanczos algorithm~\cite{nguyen2012improving,rocca2008turbo}. The projective dielectric eigenpotentials (PDEP) technique is used to compute the screened exchange integral, avoiding the need for empty bands~\cite{yu2022gpu,umari2010gw}. The Bethe-Salpeter Equation (BSE) for two-particle excitation is solved within the density matrix perturbation theory  formalism~\cite{ping2013electronic,ping2012ab}, with the dielectric matrix being computed similar to that in GW. In addition, the Yambo code~\cite{YAMBO} is used as comparison. The Godby-Needs plasmon-pole approximation (PPA) is adopted for the dynamical screening in GW~\cite{godby1989metal, oschlies1995gw}, and the BSE is solved subsequently with the GW quasiparticle energies. The details of the $\gwbse$ calculations can be found in SM Sec.~\textcolor{red}{\rn{2}}~\cite{supplementary_material}. 

Due to the multi-reference character of the $^3E$, $^1E$, and ${^1A_1}$ excited states of the NV center, a single Slater determinant description in DFT is incomplete. To overcome this problem, we use the CASSCF method~\cite{roos2005casscf} to construct the state wavefunctions as linear combinations of multiple Slater determinants. 
Each determinant describes a particular occupation of single-electron orbitals. We use the $\NVClusterM$ cluster model, which is considered in Ref.~\cite{bhandari2021multiconfigurational}. The cluster model is cut from the $\gs$ ground state geometry of $3\times3\times3$ supercell, and the C atoms that have dangling bonds are terminated H atoms with standard C-H bond length of 1.09 $\si{\angstrom}$.
In the CASSCF method, the orbital space is partitioned into subspaces of the inactive, active, and virtual orbitals. The occupation of the active orbitals is allowed to change to generate all possible spin- and symmetry-allowed determinants for a particular electronic state. A choice of a partition of the orbital space, that is the choice of a number of active orbitals and active electrons, determines the CASSCF active space. The active space of (6e, 6o) is chosen to include two molecular orbitals of the $a_1$ ($a_{1N}$ and $a_{1C}$) symmetry and two orbitals of the $e$ symmetry ($e_x$, $e_y$, $e_x^\prime$, and $e_y^\prime$), all localized near the vacancy defect. Inclusion of the $e_x^\prime$ and $e_y^\prime$ orbitals in the active space is necessary for accurate description of the correlation between electrons near the defect~\cite{bhandari2021multiconfigurational}. We use the state-average (SA) version of the CASSCF method as implemented in ORCA~\cite{neese2012orca, neese2020orca} to obtain state wavefunctions for calculations of SOC matrix elements. The ${^3A_2}$, $^3E$, ${^1A_1}$, and $^1E$ electronic states, involved in the spin polarization cycle~\cite{thiering2017ab, thiering2018theory}, are included in state-averaging. 
 Accurate prediction of the excitation energies requires correction of the CASSCF energies for the dynamic electron correlation. In this work, we use the fully internally contracted (FIC) version of the complete active space second order perturbation theory (CASPT2)~\cite{andersson1990} and $\emph{N}$-electron valence state second order perturbation theory (NEVPT2)~\cite{angeli2001} as implemented in the ORCA 5.0 program package.
 The second order Douglas-Kroll-Hess (DKH2) Hamiltonian~\cite{reiher2006dkh} and the cc-pVDZ-DK basis set~\cite{dunning1989basis, pritchard2019basis} were used to account for the scalar relativistic effects. The spin-orbit mean-field operator was used~\cite{neese2005soc}.

\section{Results and Discussion}
\subsection{ODMR Theory, Implementation, and Benchmark}

\begin{figure*}
    \centering
    \includegraphics[width=\textwidth]{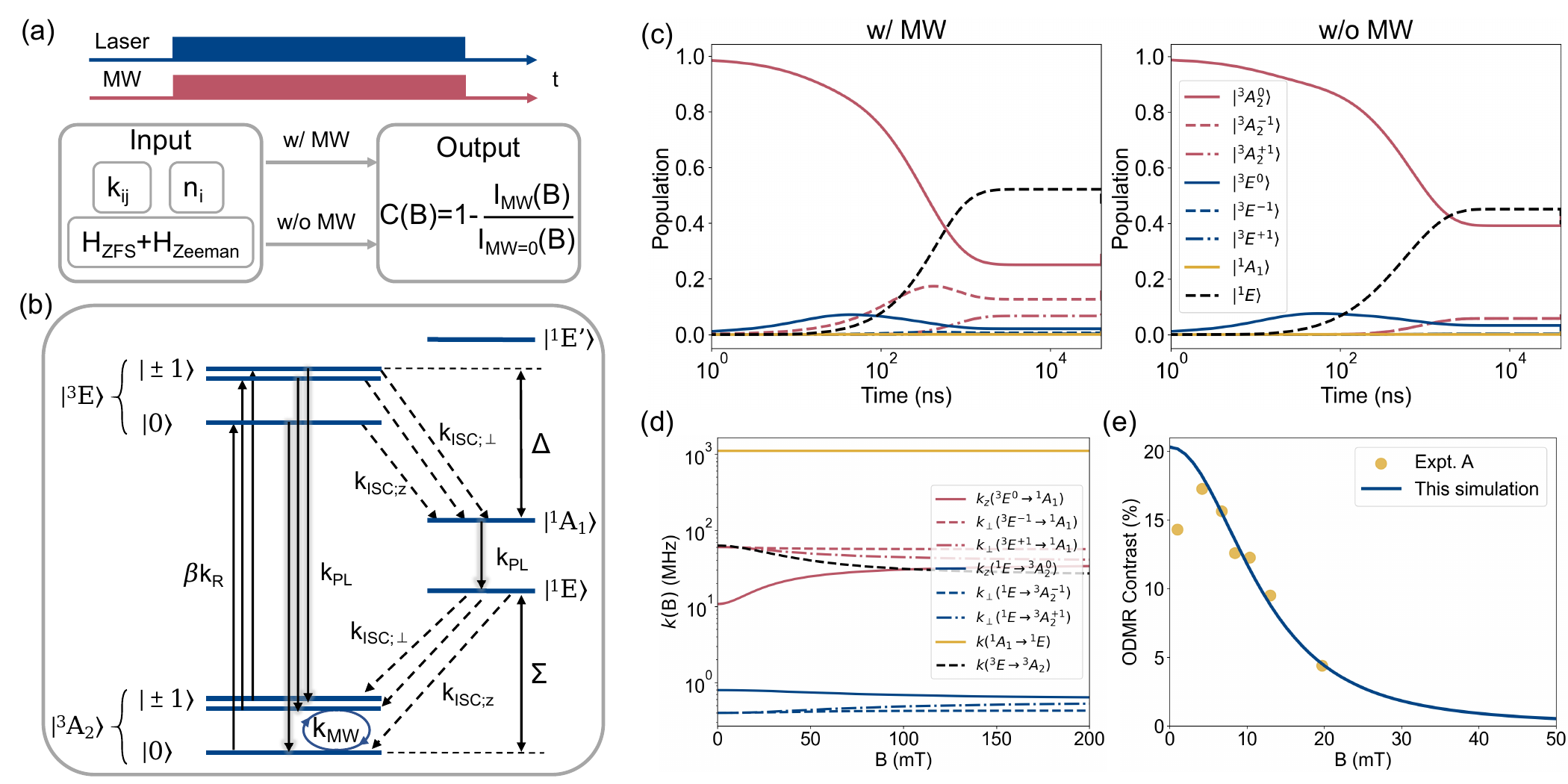}
    \caption{Numerical implementation of ODMR contrast and its benchmark with the NV center in diamond regarding the experiment~\cite{tetienne2012magnetic}, where the magnetic field is applied at an angle of 74$\si{\degree}$ with respect to the NV axis. (a) The continuous-wave (cw) ODMR and its simulation workflow with first-principles inputs. (b) Schematic diagram depicting the energy levels and excited-state kinetic processes of the NV center in diamond. $\ket{0}$, $\ket{-1}$ and $\ket{+1}$ denote the spin sublevels in the triplet states. The spin sublevels $\ket{+1}$ and $\ket{-1}$ are degenerate under zero magnetic field. (c) The time-evolved populations of the states reach a steady-state plateau with sufficiently long time, in the presence (w/MW, left panel) or absence (w/o MW, right panel) of microwave field. (d) Transition rates ($k$) for different processes vary differently with magnetic field B. (e) Simulated ODMR contrast as a function of magnetic field (solid line) and compared with the experiment (dots)~\cite{tetienne2012magnetic}. The experimental data is extracted from a figure by using WebPlotDigitizer tool~\cite{rohatgi2014webplotdigitizer}.}
    \label{fig:NV_center_ODMR_workflow}
\end{figure*}

The basic mechanism of ODMR is related to the excited-state populations of spin sublevels. Therefore, we begin with the Hamiltonian that describes the spin sublevels of triplet states as in Fig.~\ref{fig:NV_center_ODMR_workflow}(b),
\begin{align}
    H_\text{s} &= D(S_z^2 - \frac{\mathbf{S}^2}{3}) + E(S_x^2 - S_y^2) + g_e\mu_B\mathbf{B}\cdot \mathbf{S} \label{eq:zfs_zeeman_hamiltonian}\\
    \mathbf{B}\cdot \mathbf{S} &= S_z B\text{cos}\theta_B + S_xB\text{sin}\theta_B\text{cos}\phi_B + S_yB\text{sin}\theta_B\text{sin}\phi_B
\end{align}
where $S_i$ is the spin-1 operator along $i$ axis, $D$ in the first term is the axial zero-field splitting (ZFS) parameter, $E$ in the second term is the rhombic ZFS parameter, $g_e$ is the electron g-factor or gyromagnetic ratio whose value is ${\sim}2$ in the NV center, $\mu_B$ is Bohr magneton, and $\mathbf{B}$ is the external magnetic field. The ZFS lifts the spin degeneracy. The third term describes the Zeeman effect occuring under magnetic field ($\mathbf{B}$) and further leading to the mixing of the original spin sublevels. $\theta_B$ and $\phi_B$ denote the misaligned angle of the magnetic field regarding the NV axis and the azimuthal angle in the x-y plane, respectively. Since in this work, we consider the symmetry retained in the x-y plane for ZFS, the ODMR does not depend on $\phi_B$. Let $\alpha_{ij}$ be the mixing coefficient of original spin sublevels $\ket{i}$ and $\ket{j}$, then we can write the transition rates under magnetic field $k_{ij}(\mathbf{B})$ to be a function of mixing coefficients and the rates at zero magnetic field ($k^0_{pq}$),
\begin{align}
    k_{ij}(\mathbf{B})
    &= \sum_{p,q} |\alpha_{ip}(\mathbf{B})|^2 |\alpha_{jq}(\mathbf{B})|^2 k^0_{pq}. \label{eq:B_dependent_rate}
\end{align}
The spin mixing will be shown below to be the source of the magnetic field-dependent ODMR.

We then numerically implement the ODMR contrast based on the kinetic master equation~\cite{tetienne2012magnetic,dreau2011avoiding}, which integrates all transition rates and allows to simulate the dynamics of states populations, photoluminescence (PL) intensity, as well as  continuous wave (cw) or time-resolved ODMR. The model starts with the conventional definition of ODMR contrast,
\begin{align}
\begin{split}
C(\mathbf{B}) = 1 - \frac{\bar{I}(t, \mathbf{B}, k_{\mathrm{MW}})}{\bar{I}(t, \mathbf{B}, k_{\mathrm{MW}}=0)} \label{eq:odmr}
\end{split}\\
\begin{split}
\bar{I}(t,\mathbf{B}) = \eta \sum_{i \in \mathrm{ES}} \sum_{j \in \mathrm{GS}} k_{ij}(\mathbf{B}) \bar{n}_i(t,\mathbf{B}) \label{eq:steady-state_pl}
\end{split}
\end{align}
where $\bar{I}(t, \mathbf{B}, k_{\mathrm{MW}})$ is the magnetic-field dependent PL intensity at the steady state in the presence of microwave resonance (MW), which is different from the case in the absence of microwave resonance ($k_{\mathrm{MW}}=0$). Here, $k_{\mathrm{MW}}$ is the Rabi frequency of a microwave field that is applied for rotating the populations of spin sublevels, and is a parameter in the model as it scales with the amplitude of the microwave field~\cite{knight1980rabi}. $\eta$ is the collection coefficient parameter for PL intensity which depends on experimental setup but does not affect ODMR. The optical saturation parameter $\beta$ seen in Fig.~\ref{fig:NV_center_ODMR_workflow}(b) and SM Sec.~\textcolor{red}{\rn{1}}~\cite{supplementary_material} plays a role in the optical excitation. Both $k_{\mathrm{MW}}$ and $\beta$ are determined according to the experimental range~\cite{tetienne2012magnetic}. The PL intensity evolves with time depending on $n_i(t,\mathbf{B})$, the population of spin level $\ket{i}$ at time $t$, and can be solved numerically using the Euler method.

To validate the ODMR contrast implementation for triplet systems, we firstly simulate the ODMR of the prototypical system NV center, as shown in Fig.~\ref{fig:NV_center_ODMR_workflow}(b), using the experimental values of ZFS and rates~\cite{tetienne2012magnetic,kehayias2013infrared,dreau2011avoiding}. 
The purpose of this benchmark is to confirm the numerical implementation of ODMR contrast from Eq.~(\ref{eq:zfs_zeeman_hamiltonian}) to Eq.~(\ref{eq:steady-state_pl}), independent of the accuracy of electronic structure and kinetic rates, which will be discussed in detail in the next sections. 
Values for the ODMR simulation parameters are tabulated in SM Table~\textcolor{red}{S1}~\cite{supplementary_material}, and the microwave resonance is applied to drive the rotation between $\ket{\gs^0}$ and $\ket{\gs^{-1}}$. Since the simulated cw-ODMR is observed at steady state, arbitrary initial populations of the spin sublevels can be used. 

The system reaches the steady state after ${\sim}10^4$ ns, as shown in Fig.~\ref{fig:NV_center_ODMR_workflow}(c). 
From the steady-state populations, we can find that $\ket{\gs^{-1}}$ has larger population due to the Rabi oscillation in the presence of the microwave field compared to the absence of the microwave field. Because the optical excitation is mostly spin-conserving~\cite{robledo2011spin}, the population of $\ket{\tes^{-1}}$ is subsequently larger. Because the nonaxial ISC $\tes\rightarrow\fses$ is symmetrically allowed and fast, as can be seen in Fig.~\ref{fig:NV_center_ODMR_workflow}(d), it competes with the radiative recombination $\tes\rightarrow\gs$, leading to overall smaller excited-state population. Therefore, the PL intensity in the presence of the microwave field becomes smaller, and the ODMR contrast is positive as shown in Fig.~\ref{fig:NV_center_ODMR_workflow}(e). 

The ODMR contrast is a result of the difference between the axial and nonaxial ISCs, which are the  transitions between spin sublevel $\ket{0}$ and $\ket{\pm1}$ to singlet, and vice versa, eventually leading to spin polarization. The decrease of ODMR contrast with the magnetic field is a consequence of the smaller difference between axial ISC rates and nonaxial ISC rates, compared Fig.~\ref{fig:NV_center_ODMR_workflow}(d) to Fig.~\ref{fig:NV_center_ODMR_workflow}(e). The fundamental reason is the mixing of spin sublevels, which will be discussed in details in Sec.~\ref{sec:odmr}. 
To understand the magnetic field effect, first, when the magnetic field is misaligned with the NV axis, there is mixing between spin sublevels $m_s=0$ and $m_s=\pm1$. Second, the spin mixing will further mix ISC rates between different transitions
as Eq.~(\ref{eq:B_dependent_rate}).
For the NV center, the spin polarization is mainly a result of  $k_\perp(\tes\rightarrow\fses)\gg k_z(\tes\rightarrow\fses)$. When $k_\perp(\tes\rightarrow\fses)$ is similar to $k_z(\tes\rightarrow\fses)$ due to spin mixing, spin polarization is weaker, ODMR contrast is smaller. In the extreme case, where the axial ISC is equal to the nonaxial ISC, there will be no ODMR contrast because no spin polarization among $m_s$ spin sublevels.

The excellent agreement of our simulation with experiment in Fig.~\ref{fig:NV_center_ODMR_workflow}(e)
validates the numerical implementation of ODMR. 
In SM Fig.~\textcolor{red}{S2} and Fig.~\textcolor{red}{S3}~\cite{supplementary_material}, we present additional results of the benchmark regarding ODMR frequency and time-resolved PL intensity of the NV center. More importantly, we determine that the dip of ODMR at $B=0$ is a result of nonzero rhombic ZFS $E$ in Eq.~\ref{eq:zfs_zeeman_hamiltonian}, which is induced by symmetry-breaking. Detailed discussion can be found in SM Sec.~\textcolor{red}{\rn{1}}, Fig.~\textcolor{red}{S4} and Fig.~\textcolor{red}{S5}~\cite{supplementary_material}.

Besides triplet spin defects, singlet defects that have a triplet metastable state can also show ODMR signal, like ST1 defect in diamond~\cite{lee2013readout,balasubramanian2019discovery}. 
In SM Table~\textcolor{red}{S2} and Fig.~\textcolor{red}{S6}~\cite{supplementary_material}, we additionally include the detailed benchmark of the ODMR model for singlet spin defects. The benchmark for both triplet and singlet spin defects demonstrates the generality of the ODMR model.

Next, we will discuss in detail the first-principles calculations of electronic structure and excited-state kinetic rates at solid-state spin defects, specifically the NV center here. 

\subsection{Electronic Structure, Excitation Energies and Radiative Recombination of NV Center~\label{sec:electronic_structure_nv_center}}

\begin{table*}
    \caption{The excitation energies of the NV center from various theories along with experiments. The energy of $\gs$ is aligned to 0 eV, and the energies of the other states are entered with respect to $\gs$. $\Sigma$ is the excitation energy of $\sgs$ with respect to $\gs$, and $\Delta$ is the energy difference between $\tes$ and $\fses$, as illustrated in Fig.~\ref{fig:NV_center_ODMR_workflow}(b).
    }
    \begin{ruledtabular}
    \begin{tabular}{cccccc}
    Method & $\sgs$ or $\Sigma$ (eV) & $\fses$ (eV) & $\tes$ (eV) & $\Delta$ (eV) \\
    \hline
    Expt. (ZPL) & 0.325-0.411\footnotemark[1] & 1.515-1.601~\cite{acosta2010optical,rogers2008infrared,kehayias2013infrared} & 1.945~\cite{davies1976optical,goldman2015phonon,goldman2015state} & 0.344-0.430\footnotemark[2]~\cite{goldman2015phonon,goldman2015state} \\
    Expt. (absorption) & -- & 1.76\footnotemark[3]~\cite{kehayias2013infrared} & 2.180~\cite{davies1976optical} & -- \\
    CDFT (HSE, ZPL) & 0.37 & -- & 1.95 & -- \\
    $\gwbse$@PBE (absorption) & - & - & 2.40& -  \\
    SA(6)-CASSCF(6,6) (absorption) & 0.66 & 1.96 & 2.30 & 0.34 \\
    CASPT2 (absorption) & 0.55 & 1.57 & 2.22 & 0.65 \\
    CASSCF~\cite{bhandari2021multiconfigurational} & 0.25 & 1.60 & 2.14 & 0.54 \\
    QDET~\cite{sheng2022green} & 0.46 & 1.27 & 2.15 & 0.88 \\
    NEVPT2-DMET~\cite{haldar2023local} & 0.50 & 1.52 & 2.31 & 0.79 \\
    CI-CRPA~\cite{bockstedte2018ab} & 0.49 &  1.41 & 1.75\footnotemark[4] & 0.34 \\
    \end{tabular}
    \end{ruledtabular}
    \label{tab:excitation_energy}
    \footnotetext[1]{The range of $\sgs$ or $\Sigma$ was determined according to the triplet ZPL $\tes\rightarrow\gs$, the singlet ZPL $\fses\rightarrow\sgs$, and $\Delta$.}
    \footnotetext[2]{The range of $\Delta$ was extracted from the ISC rate equation in Ref.~\cite{goldman2015phonon,goldman2015state}. It is simply for reference.}
    \footnotetext[3]{This energy is obtained by adding the approximate energy (0.40 eV) of $\sgs$ to the absorption energy $\sgs \rightarrow \fses$ 1.36 eV (912 nm).}
    \footnotetext[4]{ZPL}
\end{table*}


The NV center comprises a substitutional nitrogen atom and a vacancy with one extra electron in the faced-center cubic diamond, with its local structure described by the $C_{3v}$ symmetry point group. The three-fold symmetry axis is the nitrogen-vacancy axis along the [111] direction. The ground and low-lying excited electronic states of the NV center can be determined from two one-electron orbitals, $a_1$ and $e$ ($e_x$, $e_y$), which transform as the $A_1$ and $E$ irreducible representations of the $C_{3v}$ point group, respectively. These one-electron orbitals are formed by the carbon and nitrogen dangling bonds near the vacancy defect. The $a_1^2e^2$ configuration gives rise to the ${^3A_2}$, ${^1E}$ and ${^1A_1}$ states, and $a_1e^3$ to the ${^3E}$ state. The states above are listed in the ascending energy order determined experimentally and from the group theory~\cite{rogers2008infrared, kehayias2013infrared, goldman2015phonon, thiering2018theory}. The total spin-orbit irreducible representations of these states are given in SM Table~\textcolor{red}{S3}~\cite{supplementary_material}. The derivation of the representations can be found in literature~\cite{tinkham2003group,maze2011properties,doherty2011negatively}.

Excitation energies play an important role in energy conservation in radiative (electron-photon interaction) and nonradiative (electron-phonon interaction) recombination rates. As we discussed earlier, the excited states of defects can have multi-reference character which is difficult to describe by conventional DFT. Therefore we show the calculated excitation energies from several different electronic structure methods, CDFT, $\gwbse$@PBE, CASSCF, and CASPT2 as listed in Table~\ref{tab:excitation_energy}. 
The CASPT2 method predicts the excitation energies of the ${^1A_1}$ and ${^3E}$ states in the good agreement with the experimental values and with energies obtained using the superclusters of different sizes~\cite{sheng2022green, haldar2023local}. The experimental absorption energy of the ${^1E}$ state is unknown. The CASPT2 method consistently predicts the energy of the ${^1E}$ state in the range of 0.55 to 0.60 eV, see SM Table~\textcolor{red}{S4}~\cite{supplementary_material}. This energy is in close agreement with the value of 0.50 (0.46) eV obtained with the embedding theories~\cite{sheng2022green, haldar2023local} and with the value of 0.49 eV obtained using the configuration interaction method~\cite{bockstedte2018ab}. Therefore, the CASPT2 excitation energies are in good agreement with those obtained using larger cluster models, and their use in the ODMR simulations is well-justified. We find that the NEVPT2 theory systematically overestimates energies of the ${^1E}$ and ${^3E}$ states, see SM Table~\textcolor{red}{S4}~\cite{supplementary_material}. More discussion about the excitation energies can be found in SM Sec.~\textcolor{red}{\rn{2}A}~\cite{supplementary_material}.

With the excitation energies, we employ the Fermi's golden rule to calculate radiative recombination rate. The details of calculation methods can be found in our past work~\cite{wu2019dimensionality,wu2019carrier,smart2021intersystem} and results can be found in SM Sec.~\textcolor{red}{\rn{2}C}~\cite{supplementary_material}. For the convenience of comparing our calculated results with previous experiment and calculations, here we adopt lifetime ($\tau$), which is the inverse of rate ($\tau=1/k$). Using the optical dipole moments and excitation energy from $\gwbse$@PBE calculations which includes excitonic effect accurately, we obtain $\tau_\text{R}=$10.62 ns for the $\tes\rightarrow\gs$ transition and 148 ns for the $\fses\rightarrow\sgs$ transition. 
Our calculated radiative lifetime of the $\tes\rightarrow\gs$ transition is consistent with experiment 12 ns for $\tes\rightarrow\gs$~\cite{goldman2015phonon}. The radiative lifetime of the $\fses \rightarrow \sgs$ transition has not been determined experimentally and is found to be dominated by nonradiative processes~\cite{rogers2008infrared}. Thus, 
we will mostly focus on the IC transition for $\fses \rightarrow \sgs$ as discussed in Sec.~\ref{sec:internal_conversion}.


\subsection{Spin-Orbit Coupling and Pseudo Jahn-Teller Effect~\label{sec:soc_JT}}


Spin-orbit coupling Hamiltonian can be separated into the axial/nonaxial SOC 
components ($\lambda_{z/\perp}$) with the angular momentum ladder operators ($L_\pm$ and $S_\pm$) and $z$ components ($L_z$ and $S_z$),
\begin{align}
    H_{\text{soc}} &= \frac{\lambda_\perp}{2} (L_+S_- + L_-S_+) + \lambda_z L_z S_z.
\end{align}

\begin{table*}
    \centering
    \caption{Summary of SOC matrix elements predicted by group theory, calculated at the theory level of CASSCF and from experiment. The unit is GHz. CASSCF ($C_{3v}$) SOC constants are consistent with prediction. ``CI'' stands for 
    configuration interaction that mixes $\sgs$ and ${^1E'}$. Pseudo JT effect mixes $\fses$ and $\sgs$ through electron-phonon coupling, leading to nonzero SOCs. ``--'' denotes values that do not exist or are not found.}
    \begin{ruledtabular}
    \begin{tabular}{cccccc}\\[-0.8em]
          & $\lambda_z(\tes,\tes)$ & $\lambda_\perp(\fses, \tes)$ & $\lambda_z(\fses,\tes)$ & $\lambda_\perp(\gs,\cisgs)$ & $\lambda_z(\gs,\cisgs)$ \\
        \hline\\[-0.8em]
        Group Theory $(C_{3v})$\footnotemark[1] & $\pm\lambda_z$ & $i\lambda_\perp$ & 0 & nonzero by CI & 0 \\
        SA(6)-CASSCF(6,6) ($C_{3v}$) & 14.21 & 3.96 & 0.06 & 5.22\footnotemark[2] & 0.03 \\
        \hline\\[-0.8em]
        (w/ pseudo JT) & $\lambda_z(\vibtes,\vibtes)$ & $\lambda_\perp(\vibfses, \vibtes)$ & $\lambda_z(\vibfses,\vibtes)$ & $\lambda_\perp(\vibgs,\vibsgs)$ & $\lambda_z(\vibgs,\vibsgs)$ \\
        \hline\\[-0.8em]
        Group Theory & $\pm\lambda_z$ & nonzero & nonzero by pseudo JT & nonzero by CI & nonzero by pseudo JT \\
        \makecell{Effective SOC \\ SA(6)-CASSCF(6,6)} & 14.21 & 6.75 & 0.83 & 5.05 & 7.72\\
        HSE~\cite{thiering2017ab,thiering2018theory} & 15.8\footnotemark[3] & 56.3\footnotemark[4] & -- & 18.96\footnotemark[5] & 15.8\footnotemark[6] \\
        Expt.~\cite{goldman2015phonon,batalov2009low,goldman2015state} & 5.33 & 6.4\footnotemark[7] & -- (nonzero) & -- (nonzero) & -- (nonzero) \\
    \end{tabular}
    \end{ruledtabular}
    \footnotetext[1]{The matrix elements here are expressed in terms of the reduced one-particle matrix elements, $\lambda_z = -i\hbar \mel{e \vert}{L^{A_2}}{\vert e}$ and $\lambda_\perp = -(i/\sqrt{2}) \hbar \mel{e \vert}{L^{E}}{\vert a_1}$.}
    \footnotetext[2]{The matrix element is nonzero because $\sgs$ coupled with singlet excited state ${^1E'}$ through 
    configuration interaction, and $\lambda_z(\gs,{^1E'})$ is symmetry-allowed. This coupling exists in CASSCF solutions, as can be seen in SM Table~\textcolor{red}{S9}~\cite{supplementary_material}. }
    \footnotetext[3]{$\lambda_z({^3E},{^3E})$ was calculated by $\mel{e_+}{H_\text{soc}}{e_+}$ with $\ket{e_+}=\frac{1}{\sqrt{2}}(\ket{e_x}+i\ket{e_y})$ at HSE~\cite{thiering2017ab}. The matrix element reduced to 4.8 by a reduction factor.}
    \footnotetext[4]{$\lambda_\perp(\fses,\tes)$ was calculated by $\mel{e_+}{H_\text{soc}}{a_1}$ with $\ket{e_+}=\frac{1}{\sqrt{2}}(\ket{e_x}+i\ket{e_y})$ at HSE~\cite{thiering2017ab}, assuming single-particle picture and that KS wavefunctions constructed ${^3E}$ and ${^1A_1}$ did not differ~\cite{goldman2015state}.}
    \footnotetext[5]{$\lambda_\perp(\vibgs,\vibsgs)$ was a parameter estimated by taking $\lambda_\perp(\vibgs,\vibsgs)/\lambda_z(\vibgs,\vibsgs) = 1.2$ for matching with experiment~\cite{thiering2018theory}.}
    \footnotetext[6]{$\lambda_z(\vibgs,\vibsgs)$ was calculated according to $\mel{^3E}{H_{soc}}{^3E}=\mel{e_+}{H_\text{soc}}{e_+}$~\cite{thiering2017ab,thiering2018theory}.}
    \footnotetext[7]{The matrix element is estimated by using the ratio $\lambda_\perp(\fses,\tes)/\lambda_z(\fses,\tes)=1.2$~\cite{goldman2015state}.}
    \label{tab:soc_matrix_elements}
\end{table*}

Group theory provides important information of whether a SOC matrix element 
is allowed or forbidden, from the symmetry point of view. 
And the multi-particle representation of the SOC matrix elements can be reduced to the single-particle representation by using the Wigner-Eckark theorem~\cite{doherty2011negatively}. Some SOC matrix elements were previously evaluated at the single-particle level at HSE, as shown in Table~\ref{tab:soc_matrix_elements}. However, because of their multi-reference character, the excited states cannot be properly described with a single Slater determinant~\cite{bhandari2021multiconfigurational}, as can be seen in SM Table~\textcolor{red}{S9} and Fig.~\textcolor{red}{S11}~\cite{supplementary_material}. Therefore, we perform CASSCF for the evaluation of SOC to take into account the multi-reference nature of the states, in comparison with TDDFT~\cite{de2019predicting} and previous DFT results ~\cite{thiering2017ab,thiering2018theory}.


As shown in Table~\ref{tab:soc_matrix_elements}, the SOC matrix elements from CASSCF are in accordance with the group theory prediction. Specifically, within the important SOC matrix elements listed, only $\lambda_z(\tes,\tes)$ and $\lambda_\perp(\tes,\fses)$ are symmetry-allowed. Accordingly we obtain finite values for these SOC matrix elements at CASSCF. 
We note that the nonzero $\lambda_\perp(\gs,\sgs)$ at CASSCF does not contradict to the zero value from the group theory prediction. This is because $\sgs$ couples with the higher singlet excited state ${^1E'}$ under 
configuration interaction, as also discussed in Refs~\cite{kehayias2013infrared,doherty2011negatively,maze2011properties}, resulting in $\cisgs=C\sgs+(1-C)\sgs'$ ($C$ the mixing coefficient). This leads to symmetry-allowed $\lambda_\perp(\gs,\cisgs)$.
%
%
In comparison, the SOC matrix elements from TDDFT do not agree with the group theory prediction. 
The issue is that TDDFT does not describe the multi-reference state correctly, as can be seen in SM Table~\textcolor{red}{S10}~\cite{supplementary_material}.
%
An apparent error is that the symmetry of the wavefunction is not preserved, as can be seen in SM Fig.~\textcolor{red}{S12}~\cite{supplementary_material}. 

However, experiments show allowed axial ISC for $\tes\rightarrow\fses$ and $\sgs\rightarrow\gs$~\cite{goldman2015phonon,tetienne2012magnetic,gupta2016efficient,robledo2011spin}. This is contradictory to the current group theory prediction that these axial ISC transitions are forbidden by zero $\lambda_z(\tes,\fses)$ and $\lambda_z(\gs,\sgs)$. It
implies that an additional mechanism may have modified the symmetry of wavefunctions. Previous studies attribute it to the pseudo JT effect, which particularly couples nondegenerate electronic states through electron-phonon coupling~\cite{bersuker2017jahn,bersuker_2006,thiering2018theory,jin2022vibrationally,razinkovas2021vibrational},
\begin{align}
    H &= \underbrace{\begin{pmatrix}
        \Lambda & 0 & 0\\
        0 & 0 & 0 \\
        0 & 0 & 0
    \end{pmatrix} }_{H_\text{e}}
    + \underbrace{\frac{K}{2} (Q_x^2 + Q_y^2) }_{H_\text{osc}}
    + \underbrace{ \begin{pmatrix}
        0 & FQ_x & FQ_y\\
        FQ_x & 0 & 0 \\
        FQ_y & 0 & 0
    \end{pmatrix} }_{H_\text{PJT}}\\
    F &= \mel{^1A_1}{\frac{\partial V}{\partial Q_x}}{{^1E_x}} = \mel{^1A_1}{\frac{\partial V}{\partial Q_y}}{{^1E_y}}
\end{align}
Here, $\Lambda$ is the energy gap between $\ket{^1A_1}$ and $\ket{^1E}$ in the electronic Hamiltonian $H_\text{e}$ . $K$ is the elastic vibronic constant in the harmonic oscillator Hamiltonian $H_\text{osc}$. $F$ is the linear vibronic coupling constant in the pseudo JT Hamiltonian $H_\text{PJT}$. $\ket{^1A_1}$, $\ket{^1E_x}$ and $\ket{^1E_y}$ form the basis of the Hamiltonian $H$. $Q_x$ and $Q_y$ are the nuclear coordinate transform as $x$ and $y$, respectively.
Under the pseudo JT distortion, $\fses$ is coupled with $\sgs$ with the assistance of the JT-active $e$ phonon, denoted as $(A_1 + E)\otimes e$. The coupling of electronic states results in the vibronic states $\vibsgs$ and $\vibfses$ as the linear combination of $\sgs$ and $\fses$, 
\begin{align}
\begin{split}
\ket{^1\widetilde{E}_{\pm}} 
&= \sum_{i=1}^\infty \Bigg[ c_i \ket{^1\bar{E}_{\pm}} \otimes \ket{\chi_i^{A_1}} + d_i \ket{^1A_1} \otimes \ket{\chi_i^{E_\pm}} \\
&\quad+ f_i \ket{^1\bar{E}_{\pm}} \otimes \ket{\chi_i^{E_\pm}} + g_i \ket{^1\bar{E}_{\mp}} \otimes \ket{\chi_i^{A_2}} \Bigg]
\end{split}~\label{eq:1E_pz_PJT}\\
\begin{split}
\ket{^1\widetilde{A}_{1}} &= \sum_{i=1}^\infty \Bigg[ c'_i \ket{^1A_1} \otimes \ket{\chi_i^{A_1}} \\
&\quad+ \frac{d'_i}{\sqrt{2}} \bigg(  \ket{^1\bar{E}_{+}} \otimes \ket{\chi_i^{E_-}} + \ket{^1\bar{E}_{-}} \otimes \ket{\chi_i^{E_+}} \bigg) \Bigg]
\end{split}~\label{eq:1A1_pz_PJT}
\end{align}
where $\ket{\chi_i^\Gamma}$ is the $i$-th phonon wavefunction transforming as irreducible representation $\Gamma$, and $c_i$, $d_i$, $f_i$, $g_i$, $c'_i$ and $d'_i$ are the amplitude of vibronic states. The states coupling give rise to a significant change in the SOC matrix elements as listed below:
%
\begin{align}
\begin{split}
\lambda_z(\vibfses,\vibtes) &= d'_{eff} \mel{\cisgs}{H_\text{soc}}{\tes}_z
\end{split}~\label{eq:axial_SOC_1A1_3E}\\
\begin{split}
\lambda_\perp(\vibfses,\vibtes) &= c'_{eff} \mel{\fses}{H_\text{soc}}{\tes}_\perp + d'_{eff} \mel{\cisgs}{H_\text{soc}}{\tes}_\perp
\end{split}~\label{eq:nonaxial_SOC_1A1_3E}\\
\begin{split}
\lambda_z(\vibgs,\vibsgs) &= d_{eff} \mel{\gs}{H_\text{soc}}{\fses}_z
\end{split}~\label{eq:axial_SOC_3A2_1E}\\
\begin{split}
\lambda_\perp(\vibgs,\vibsgs) &= (c_{eff} + f_{eff} ) \mel{\gs}{H_\text{soc}}{\cisgs}_\perp
\end{split}~\label{eq:nonaxial_SOC_3A2_1E}
\end{align}
where $c'_{eff}$, $d'_{eff}$, $c_{eff}$, $d_{eff}$, $f_{eff}$ are the normalized effective state mixing coefficients, e.g. $|c'_{eff}|^2 + |d'_{eff}|^2 = 1$. 

We complete the derivation for all effective SOC matrix elements of the NV center, in complement to Ref.~\cite{thiering2018theory} which only provides $\lambda_{\perp/z}(\vibgs,\vibsgs)$. The derivation details can be found in SM Sec.~\textcolor{red}{\rn{4}A}~\cite{supplementary_material}. 
The pseudo JT effect results in the finite values of $\lambda_z(\vibfses,\vibtes)$ and $\lambda_z(\vibgs,\vibsgs)$, and a modification to $\lambda_\perp(\vibfses,\vibtes)$ and $\lambda_\perp(\vibgs,\vibsgs)$, as can be seen in Table~\ref{tab:soc_matrix_elements}.
The resulting effective SOC shows much better agreement with experiments, in contrast to other calculations by DFT or TDDFT. 
As will be shown in Sec.~\ref{sec:eph_isc}, the nonzero SOC matrix elements lead to allowed ISC, consistent with experimental observations.

Finally, another approach based on the Taylor expansion of SOC matrix element in terms of coupling with phonon vibration 
is commonly used for the study of ISC in molecules~\cite{lawetz1972theory,tatchen2007intersystem}. We examine this approach by computing its first-order derivative of SOC with respect to nuclear coordinate change. The result indicates that this approach does not apply to the NV center under the 1D effective phonon approximation. 
More details can be seen in the SM Sec.~\textcolor{red}{\rn{4}C}~\cite{supplementary_material}. 

\begin{table*}
    \centering
    \caption{Electron-phonon coupling and rates of ISC $\vibtes\rightarrow\vibfses$ and $\vibsgs\rightarrow\vibgs$ at 300 K. The degree of geometry degeneracy $g$ is specified as 3. $\Delta Q$ is the nuclear coordinate change between the initial and final states. $\hbar\omega_i$ and $\hbar\omega_f$ represent the phonon energy of the initial and final states, respectively. $S_f$ denotes the 1D Huang-Rhys (HR) factor of the final state, which is a measure of e-ph coupling strength. $\widetilde{X}_{if}$ is the phonon term. In these calculations, the JT-distorted geometries of $C_s$ symmetry are used for the initial states $\vibtes$ and $\vibsgs$, and the geometry of $C_{3v}$ symmetry of $\vibgs$ is used for the final states $\vibfses$ and $\vibgs$, considering the fact that the geometry and phonon modes of $\vibfses$ are similar to those of $\vibgs$~\cite{jin2022vibrationally,kehayias2013infrared}. 
    } 
    \begin{ruledtabular}
    \begin{tabular}{ccccccccccccc}
        & Transition & $\Delta Q$ & $\hbar\omega_i$ & $\hbar\omega_f$ & $S_f$ & $\widetilde{X}_{if}$ & $k_{\perp}$ & $k_z$\\
        & & (amu$^{1/2}\si{\angstrom}$) & (meV) & (meV) &  & ($\mathrm{eV}^{-1}$) & (MHz) & (MHz)\\
        \hline\\[-0.8em]
        Expt. & \multirow{3}{*}{$\vibtes \rightarrow \vibfses$} & 0.64\footnotemark[1] & -- & 71~\cite{kehayias2013infrared} & 3.49~\cite{kehayias2013infrared} & -- & 24.3~\cite{goldman2015phonon} & $<0.62$~\cite{goldman2015phonon} \\
        Calc. &  & 0.65 & 73.74 & 67.75 & 3.38 & 1.34 & 29.86 & 0.46 \\
        Calc.~\cite{thiering2017ab,thiering2018theory} &  & -- & 77.6 & $66-91.8$ & 2.61\footnotemark[2] & -- & 243 & -- \\
        \hline\\[-0.8em]
        Expt. & \multirow{6}{*}{$\vibsgs \rightarrow \vibgs$} & 0.34\footnotemark[1] & -- & 64~\cite{kehayias2013infrared} & 0.9\footnotemark[3]~\cite{kehayias2013infrared} & -- & 2.61~\cite{robledo2011spin} & 3.0~\cite{robledo2011spin} \\
        \makecell{Calc. \\ ($\widetilde{X}_{if}$ at HSE)} &  & 0.23 & 75.85 & 71.04 & 0.45 & $1.17\times10^{-4}$ & $1.47\times10^{-3}$ & $3.47\times10^{-3}$ \\
        \makecell{Calc.\\($\widetilde{X}_{if}$ at spin-flip TDDFT)} &  & 0.41~\cite{jin2022vibrationally} & 48.81 & 69.88 & 1.46 & 0.20 & 2.49 & 5.89 \\
        Calc.~\cite{thiering2018theory} &  & -- & -- & 66 & 0.07 & -- & 0.90\footnotemark[4] & 4.95\footnotemark[4] \\
    \end{tabular}
    \end{ruledtabular}
    \label{tab:e-ph_ISC}
    \footnotetext[1]{$\Delta Q$ is estimated by the Huang-Rhys factor and phonon energy, $S=\omega\Delta Q^2/2\hbar$, under one-dimensional effective phonon approximation.}
    \footnotetext[2]{$S=2.61$ when approximating $e_xe_x$ singlet determinant for $\vibfses$, and $S=3.11$ when using $\gs$ geometry for $\vibfses$.}
    \footnotetext[3]{This HR factor is for $\vibfses\rightarrow\vibsgs$. Considering the similarity of geometry and potential surfaces between $\vibgs$ and $\vibfses$, we put it here for comparison with our calculations.
    }
    \footnotetext[4]{The corresponding SOC can be found in Table~\ref{tab:soc_matrix_elements}.}
\end{table*}

\subsection{Electron-Phonon Coupling and ISC~\label{sec:eph_isc}}

The ISC rate between electronic states with different spin multiplicities can be calculated as
\begin{align}
    k_\text{ISC} &= \frac{2\pi}{\hbar}g\sum_{n,m}p_{in}|\mel{fm}{H_\text{soc}}{in}|^2\delta(E_{fm}-E_{in}).~\label{eq:ISC_FGR}
\end{align}
Here, we make the 1D effective phonon approximation~\cite{nonrad_alkauskas2014first,wu2019carrier,smart2021intersystem} for the ISC rate. Compared to the full-phonon method used in Ref.~\cite{thiering2017ab,thiering2018theory}, an advantage of this 1D effective phonon method is that it enables the use of different values for the phonon energy of 
initial state ($\hbar\omega_i$) and the one of final state ($\hbar\omega_f$), and enables finite temperature occupation for both states. 
As already discussed in Sec.~\ref{sec:soc_JT} and SM Sec.~\textcolor{red}{\rn{4}A}~\cite{supplementary_material}, the SOC matrix element $\mel{fm}{H_\text{soc}}{in}$ that couples the initial vibronic state $\ket{in}$ and final vibronic state $\ket{fm}$ can be separated into the effective SOC and effective phonon overlap. Therefore, the ISC rate equation is expressed as
\begin{align}
    k_\text{ISC} &= \frac{2\pi}{\hbar}g\lambda^2\widetilde{X}_{if}~\label{eq:derived_ISC}\\
    \lambda &= \mel{\psi_f}{H_{\text{soc}}}{\psi_i} \\
    \widetilde{X}_{if} &= \sum_{n,m}p_{in}(T) |\braket{\phi_m}{\phi_n}|^2 \delta(m\hbar\omega_f-n\hbar\omega_i+\Delta E_{if}) 
\end{align}
where $g$ is the degree of degeneracy on equivalent structural configuration, $\lambda$ is the effective SOC matrix element, and $\widetilde{X}_{if}$ is the temperature-dependent phonon term representing the phonon contribution. $\ket{\psi}$ is the linear combination of possible electronic states after considering pseudo JT effect, and $\ket{\phi}$ denotes the phonon wavefunction under the harmonic oscillator approximation. $\lambda$ has been discussed in detail in Sec.~\ref{sec:soc_JT}. Therefore, our primary focus in this section will be on the phonon term $\widetilde{X}_{if}$. We will show that dynamical JT effect is important to include for electron-phonon coupling in the nonradiative intersystem-crossing transitions. Otherwise the transition rates can be order of magnitude smaller. 

\begin{figure}[h!]
    \centering
    \includegraphics[width=0.42\textwidth]{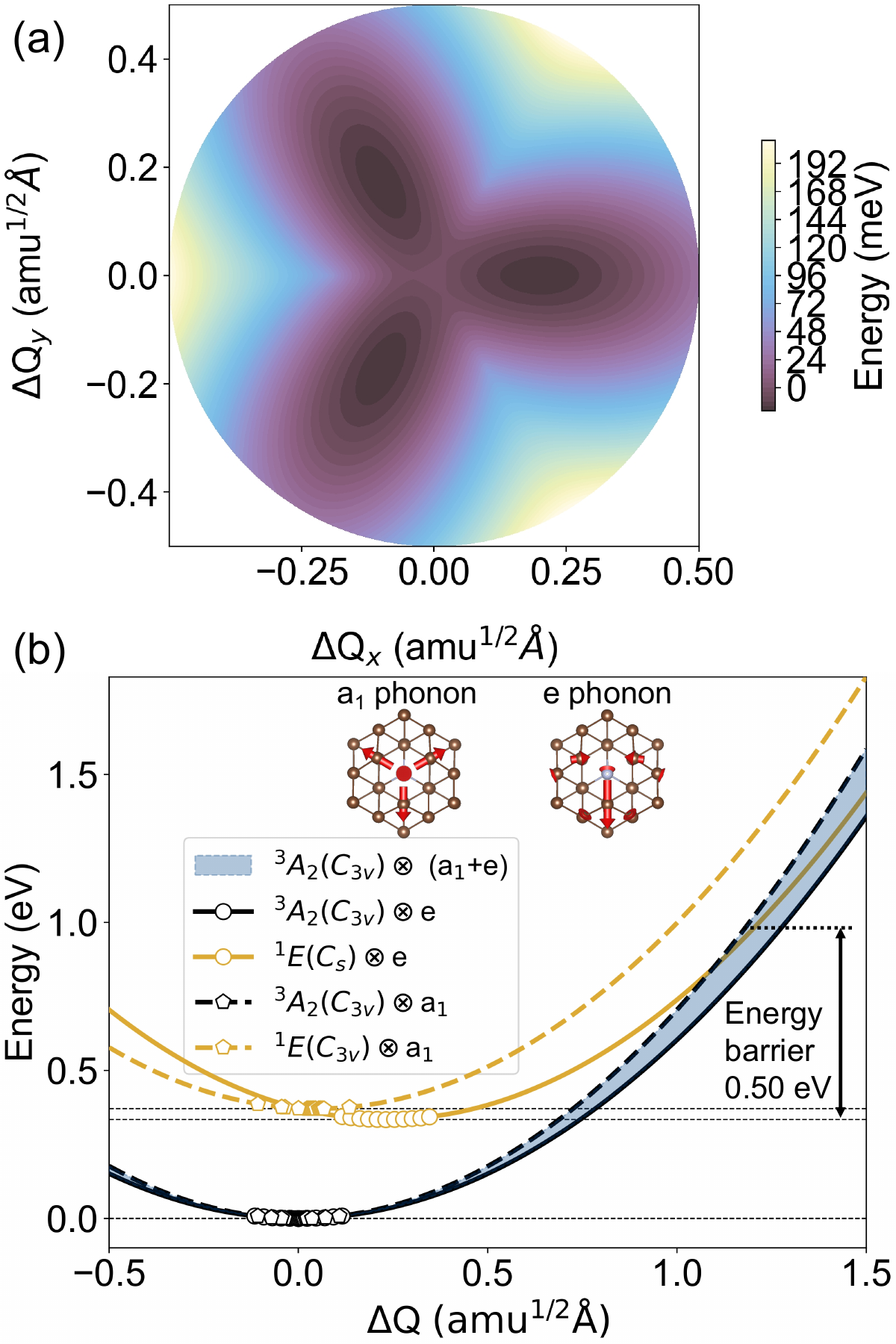}
    \caption{(a) Adiabatic potential energy surface (APES) of $\sgs$ of the NV center with the dynamical JT effect. $\Delta Q$ denotes the nuclear coordinate change, the mass weighted total nuclear displacement. $x$ and $y$ represent the displacement directions due to the degenerate $e_x$ and $e_y$ phonons, respectively. (b) The configuration coordinate diagram for the $\vibsgs\rightarrow\vibgs$ ISC. The ISC starts from $\vibsgs$, whose geometry at local energy minima can have $C_{3v}$ and $C_s$ symmetries, and ends at $\vibgs$ of $C_{3v}$ symmetry. The solid line represents the potential energy curve created by electron coupling with $e$ phonon, and the dashed line represents that due to electron coupling with $a_1$ phonon. The shaded area between the solid and dashed lines includes electron coupling with the mix of $a_1$ and $e$ phonons. The energy barrier is 0.50 eV. The inset is the visualization of effective $a_1$ and $e$ phonons, whose arrows represent the vibration amplitude larger than the threshold 0.005$~\mathrm{amu}^{1/2}\si{\angstrom}$ and 0.03$~\mathrm{amu}^{1/2}\si{\angstrom}$, respectively.}
    \label{fig:DJT_1E_APES_ISC}
\end{figure}

In Fig.~\ref{fig:DJT_1E_APES_ISC}(a), we show the lower branch adiabatic potential energy surface (APES) of $\sgs$ by fitting the Hamiltonian in Eq.~(\ref{eq:DJT})~\cite{bersuker_2006} to the calculated potential energy curves,
\begin{align}
\begin{split}
    H &= \underbrace{\frac{K}{2} (Q_x^2 + Q_y^2)}_{H_\text{osc}} \\&\quad+ \underbrace{F(Q_x\sigma_z - Q_y\sigma_x) + G[(Q_x^2-Q_y^2)\sigma_z + 2Q_xQ_y\sigma_x]}_{H_\text{DJT}}.~\label{eq:DJT}
\end{split}
\end{align}
Here $K$, $F$ and $G$ are elastic, linear and quadratic vibronic constants, respectively. $\sigma_z$ and $\sigma_x$ are Pauli's matrices. $Q_x$ and $Q_y$ denote nuclear coordinate transform as $x$ and $y$, respectively. The APES appears as a tricorn Mexican hat under the $E\otimes e$ JT distortion, which splits the doubly degenerate electronic states via the coupling with $e$ phonon. The symmetry of the geometry is $C_{3v}$ at the conical intersection, and $C_s$ at the three equivalent geometries of energy minima. The three-fold degeneracy of the geometry indicates that $g$ should be 3 in the ISC calculation. We find that the three energy minima are separated by a minor energy barrier ($\delta=2F^2|G|/(K^2-4G^2)$) at ${\sim}30$ meV. Because of the small energy barrier, the system of the vibronic ground state may be delocalized and tend to undergo a hindered internal rotation among the energy minima like $\tes$~\cite{abtew2011dynamic}, which is the so-called ``dynamical JT effect". 

The dynamical JT effect distorts the geometries from $C_{3v}$ to lower symmetry. Then the geometries of $C_s$ symmetry become the starting points for the system to relax from the initial states $\vibsgs$ ($\vibtes$) to the final state $\vibgs$ ($\vibfses$) through ISC. The nonradiative process is depicted in Fig.~\ref{fig:DJT_1E_APES_ISC}(b), the configuration coordinate diagram of $\vibsgs\rightarrow\vibgs$ as an example. In the nonradiative process, the JT-active $e$ phonon breaks the symmetry of $\sgs$ and alters the potential energy curve of $\sgs$. This results in a small energy barrier between the initial and final electronic states, which is easy to overcome. Finally, the small energy barrier consequences in the fast nonradiative relaxation~\cite{brawand2015surface, wu2019carrier}. In contrast, if not considering the dynamical JT effect, only the totally symmetric $a_1$ type phonon would participate in the ISC, and the transition rate would be orders of magnitude smaller due to the large energy barrier between the initial and final electronic states. The dramatic difference between $C_{3v}$-symmetry geometry and $C_s$-symmetry geometry as starting point indicates that dynamical JT effect plays an essential role in the nonradiative processes.

By listing the ISC calculation details in Table~\ref{tab:e-ph_ISC}, we first find $k_{\perp}(\vibtes\rightarrow\vibfses)$ shows good agreement with experimental values. And 
$k_z(\vibtes\rightarrow\vibfses)$ becomes allowed in the first order due to the finite SOC by the pseudo JT effect, consistent with the nonzero ISC in experiment~\cite{goldman2015phonon,tetienne2012magnetic,gupta2016efficient,robledo2011spin}. On the other hand, the $k_{\perp/z}(\vibsgs\rightarrow\vibgs)$ ISC rates are three orders of magnitude smaller than experiment, but the ratio $k_z(\vibsgs\rightarrow\vibgs)/k_\perp(\vibsgs\rightarrow\vibgs)=$ 2.37 is similar to the experimental value $1.15\pm0.05$ or $1.6\pm0.4$~\cite{robledo2011spin} (the ratio between $k_z$ and $k_{\perp}$ is more critical for spin polarization). The underestimation of the rates can be originated from the underestimated electron-phonon coupling related to $\widetilde{X}_{if}$, the phonon contribution to ISC. 
Recent spin-flip TDDFT calculation~\cite{jin2022vibrationally} predicts accurate PL lineshape for the $\sgs\rightarrow\fses$ transition, which indicates accurate electron-phonon coupling description despite the lack of double excitation. We therefore use spin-flip TDDFT calculation to correct the phonon term $\widetilde{X}_{if}$ of this transition.
It can be found that the phonon term is underestimated by orders of magnitude at HSE compared to spin-flip TDDFT.

In fact, the underestimation of electron-phonon coupling can be traced back to the JT effects, including the dynamical JT effect and pseudo JT effect. Our HSE calculation captures the dynamical JT effect, giving $\Delta Q = 0.23$ amu$^{1/2}\si{\angstrom}$. This value is similar to the value 0.26 amu$^{1/2}\si{\angstrom}$ obtained with spin-flip TDDFT~\cite{jin2022vibrationally}. However, due to the mean-field approximation and the absence of spin-flip process in HSE, the pseudo JT effect is relatively weak in the $\sgs$ state. This results in the lack of the additional distortion of $\sgs$ by the pseudo JT effect in our HSE calculation. If using spin-flip TDDFT, $\Delta Q$ is enhanced to 0.42 amu$^{1/2}\si{\angstrom}$, which is closer to the experiment 0.34 amu$^{1/2}\si{\angstrom}$ that is estimated by the experiment phonon energy and Huang-Rhys factor. With the spin-flip TDDFT data, we obtain $k_{\perp/z}(\vibsgs\rightarrow\vibgs)$ in good agreement with the experimental observation, as listed in Table~\ref{tab:e-ph_ISC}.

\subsection{Internal Conversion~\label{sec:internal_conversion}}
The IC represents spin-conserving phonon-assisted nonradiative transition. Under the static coupling approximation and one-dimensional effective phonon approximation, the equation of nonradiative transition rates is expressed as~\cite{nonrad_alkauskas2014first,wu2019carrier}

\begin{align}
\begin{split}
    k_\text{IC} &= \frac{2\pi}{\hbar}g|W_{if}|^2X_{if}(T) \label{eq:nonrad_wif_xif}\\
\end{split}\\
\begin{split}
    W_{if} &=(\varepsilon_f-\varepsilon_i) \braket{\psi_i(\mathbf{r, R})}{\frac{\partial \psi_f(\mathbf{r, R})}{\partial Q}} \bigg|_{\mathbf{R=R_a}} \label{eq:wif}\\
\end{split}\\
\begin{split}
    X_{if} &= \sum_{n,m}p_{in}(T)\Big|\mel{\phi_{fm}(\mathbf{R})}{Q-Q_a}{\phi_{in}(\mathbf{R})}\Big|^2\\ &\quad\times\delta(m\hbar\omega_f-n\hbar\omega_i+\Delta E_{if}). \label{eq:xif}
\end{split}
\end{align}
This equation is similar to that of ISC except that the electronic term $W_{if}$ replaces SOC $\lambda$ with single particle wavefunctions $\psi_i$ to approximate many-electron wavefunctions, and that the phonon term becomes $X_{if}$ with additional expectation values of nuclear coordinate change ($Q-Q_a$).

From our calculations reported in Table~\ref{tab:internal_conversion_1A1_1E}, $\tes\rightarrow\gs$ shows long IC lifetime (1.20 s) because a great number of emitted phonons is needed to fulfill the energy conservation between $\tes$ and $\gs$. Therefore, this spin-conserving transition process is dominated by the radiative recombination.
%
On the other hand, the $\fses\rightarrow\sgs$ transition is likely to be nonradiative-dominant according to past experiments~\cite{rogers2008infrared}, but does not appear in the HSE calculation where we obtain long IC lifetime. This can be a consequence of inaccurate description of the electron-phonon coupling of multi-reference states $\fses$ and $\sgs$, currently calculated at DFT.

Similar to ISC, we also correct 
the phonon term $X_{if}$ of $\sgs\rightarrow\fses$ transition, which is the primarily underestimated component, by using the potential energy surfaces of $\fses$ and $\sgs$ at spin-flip TDDFT@DDH. The improved electron-phonon coupling enhances $X_{if}$ and the IC rate  by several orders of magnitude. The enhanced IC rate is in good agreement with the experimental observation~\cite{acosta2010optical}. More details can be found in SM Sec.~\textcolor{red}{\rn{5}D}~\cite{supplementary_material}.

Without correcting the IC rate, we find that the  PL rate ($k_\text{PL}=k_\text{R}+k_\text{IC}$) of $\fses\rightarrow\sgs$ is underestimated, but still allows us to obtain qualitatively correct ODMR contrast. When we consider the corrected $k_\text{IC}(\vibfses\rightarrow\vibsgs)$ by using spin-flip TDDFT data, this transition becomes dominant by nonradiative recombination, and ODMR contrast is in better agreement with experiments. More details can be seen in Sec.~\ref{sec:odmr}.

\begin{table*}
    \centering
    \caption{Calculation details of internal conversion of the NV center at 300 K. The degree of geometry degeneracy $g$ is specified as 3. The ZPL is the energy change in the complete nonradiative process. $\Delta Q$ denotes the nuclear coordinate change between the initial and final states. $S_f=\omega \Delta Q^2/2\hbar$ is the phonon energy of the final state. $S_f$ denotes the Huang-Rhys factor under 1D effective phonon approximation. $W_{if}$ and $X_{if}$ are the electronic and phonon terms, respectively. 
    } 
    \begin{ruledtabular}
    \begin{tabular}{cccccccccc}
        Transition & ZPL & $\Delta Q$ & $\hbar\omega_f$ & $S_f$ & $W_{if}$ &  $X_{if}$ & $\tau_{\mathrm{IC}}$ & $k_{\mathrm{IC}}$ \\
         & (eV)& (amu$^{1/2}\si{\angstrom}$) & (meV) & & (eV/(amu$^{1/2}\si{\angstrom}$)) & (amu$\cdot\si{\angstrom}^2$/eV) & & (MHz) \\
        \hline\\[-0.8em]
        \makecell{$\tes\rightarrow\gs$} & 1.97 & 0.65 & 67.75 & 3.38 & 16.59 & $1.06\times10^{-19}$ & 1.20 s & $8.35\times10^{-7}$ \\
        \makecell{$\fses\rightarrow\sgs$ \\($X_{if}$ at HSE)}  & 1.13 & 0.24 & 75.69 & 0.51 & $1.05\times10^{-2}$ &  $4.28\times10^{-11}$ & 7.35 ms & $1.36\times10^{-4}$\\
        \makecell{$\vibfses\rightarrow\vibsgs$ \\($X_{if}$ at spin-flip TDDFT)} & 1.13 & 0.42~\cite{jin2022vibrationally} & 87.33 & 1.82 & $1.05\times10^{-2}$ & $1.52\times10^{-4}$~\footnotemark[1] & 2.07 ns & $4.82\times10^{2}$ \\
        \makecell{$\fses\rightarrow\sgs$ (Expt.)} & 1.19 & 0.31-0.35\footnotemark[2] & -- & 0.9~\cite{kehayias2013infrared} & -- & -- & 0.9 ns\footnotemark[3]~\cite{acosta2010optical} & $1.11\times10^{3}$~\footnotemark[3]~\cite{acosta2010optical}\\
    \end{tabular}
    \end{ruledtabular}
    \footnotetext[2]{$\Delta Q$ is estimated by the Huang-Rhys factor and phonon energy in the range 63-76.4 meV, $S=\omega\Delta Q^2/2\hbar$, under one-dimensional effective phonon approximation.}
    \footnotetext[3]{These are the PL lifetime and rate of $\fses$. Since the transition is claimed to be dominated by nonradiative processes~\cite{rogers2008infrared}, they are approximately IC lifetime and rate.}
    \label{tab:internal_conversion_1A1_1E}
\end{table*}

\subsection{Angle-Dependent and Magnetic-Field Dependent ODMR~\label{sec:odmr}}
The rates of radiative recombination, internal conversion and intersystem crossing obtained from the calculations above set the prerequisite for the simulation of ODMR contrast (spin-dependent PL contrast). Additionally, the ZFS
of triplet ground state and excited state are entered to account for their spin sublevels. Our calculated ZFS of $\gs$ is $D=3.03$ GHz 
by the first-principles method explained in SM Sec.~\textcolor{red}{\rn{6}A}~\cite{supplementary_material}, similar to the experimental value $D=2.87$ GHz~\cite{smart2021intersystem,tetienne2012magnetic,neumann2009excited}. We currently use the experimental value of $D=1.42$ GHz for excited triplet state $\tes$~\cite{neumann2009excited} given methods for accurate prediction of excited state ZFS remain to be developed. The optical saturation parameter $\beta$ and Rabi frequency $k_\text{MW}$ are entered as parameters into the model, and their values are selected within the experimental range~\cite{dreau2011avoiding}. cw-ODMR, which is evaluated at the steady state when the populations no longer change as a function of time, is independent on the initial spin state. 
We then apply an oscillatory microwave field to drive the population between $\ket{-1}$ and $\ket{0}$ spin sublevels in $\gs$.

\begin{figure}[h!]
    \centering
    \includegraphics[width=0.45\textwidth]{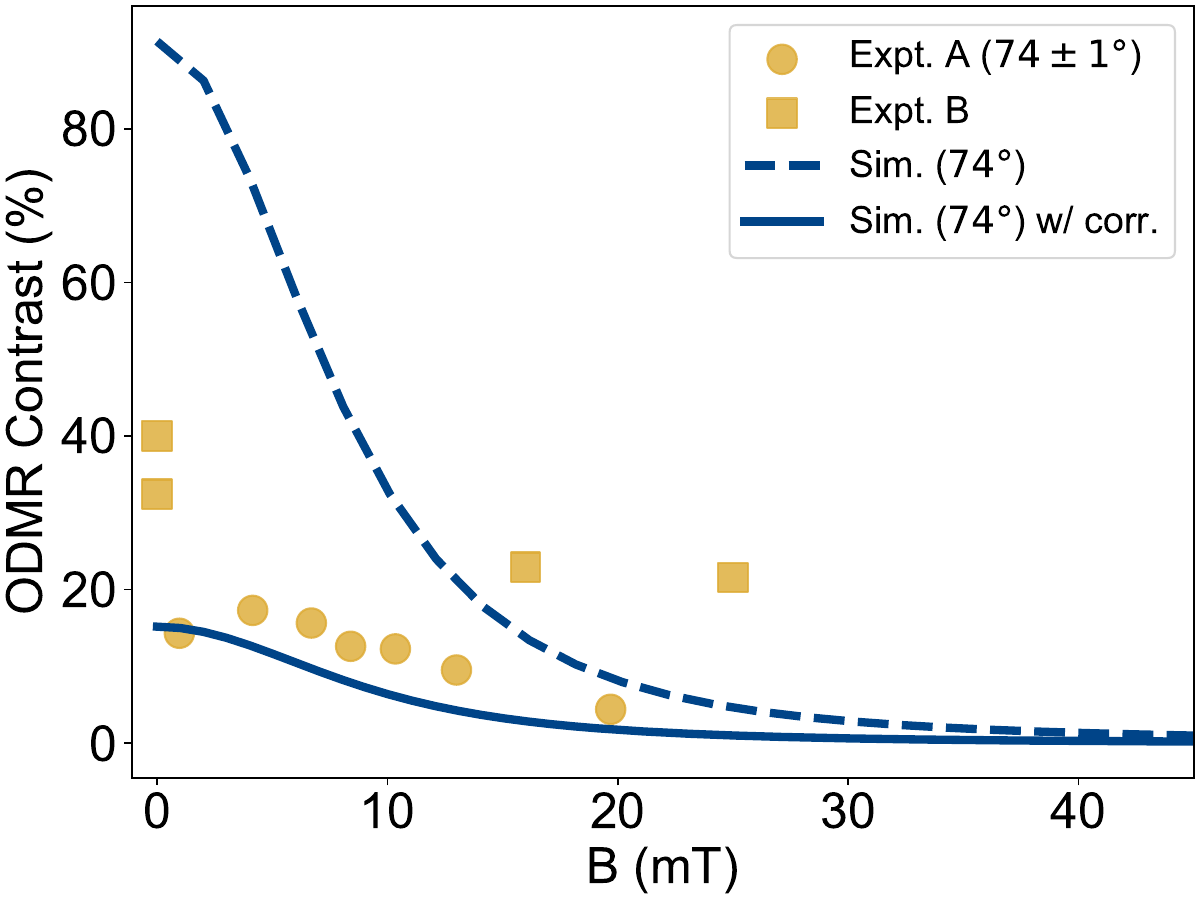}
    \caption{ODMR contrast from the simulation using the calculated rates compared with Expt. A~\cite{tetienne2012magnetic} and Expt. B~\cite{janitz2022diamond}. The angle refers to the angle of magnetic field with respect to the NV axis along [111]. Expt. A shows an angle of $74\pm1\si{\degree}$. The angle of magnetic field in Expt. B is unknown. The dashed blue curve is the ODMR simulation using all the rates from calculations using HSE phonon. The solid blue curve is the simulation  with calculated rates using the phonon term calculated with spin-flip TDDFT as discussed in Sec.~\ref{sec:eph_isc} and Sec.~\ref{sec:internal_conversion}.
    }
    \label{fig:ODMR_vs_theta_B}
\end{figure}

Using the input from our first-principles calculations, we plot the simulated ODMR contrast against magnetic field in comparison with the experiments~\cite{tetienne2012magnetic,janitz2022diamond} in Fig.~\ref{fig:ODMR_vs_theta_B}. The simulated ODMR contrast decreases with increasing the magnetic field, consistent with the trend shown by the experiments (yellow dots and squares). Such decrease of ODMR contrast is originated from the mixing of the spin sublevels under magnetic field, as can be seen in SM Fig.~\textcolor{red}{S21}~\cite{supplementary_material}. The mixing of the spin sublevels leads to smaller contrast of the axial and nonaxial ISC rates, especially $k_{\perp/z}(\vibtes\rightarrow\vibfses)$. Consequently, the spin polarization is less pronounced, manifesting as reduced ODMR contrast. 

Our simulated ODMR contrast is overestimated compared to experiments. This is because of the underestimated ISC rates $k_{\perp/z}(\vibsgs\rightarrow\vibgs)$ and IC rate $k_\text{IC}(\vibfses\rightarrow\vibsgs)$, with detailed explanation in earlier sections. Under continuous optical excitation, the system is easier to populate $\vibsgs$ and $\vibfses$ through the channel $\vibtes^{\pm1}\rightarrow\vibfses\rightarrow\vibsgs$ when there is an applied microwave field driving populations from spin sublevel $\ket{0}$ to $\ket{-1}$. When $k_{\perp/z}(\vibsgs\rightarrow\vibgs)$ and $k_\text{IC}(\vibfses\rightarrow\vibsgs)$ are underestimated, the population accumulates at the singlet states $\vibfses$ and $\vibsgs$. This results in the enhanced PL intensity contrast between the two situations, namely in the presence and absence of a microwave field. With the calculated rates using spin-flip TDDFT data discussed in Sec.~\ref{sec:internal_conversion} and Sec.~\ref{sec:eph_isc}, we obtain ODMR contrast slightly underestimated but in better agreement with the experiments. 

\begin{figure}[h!]
    \centering
    \includegraphics[width=0.45\textwidth]{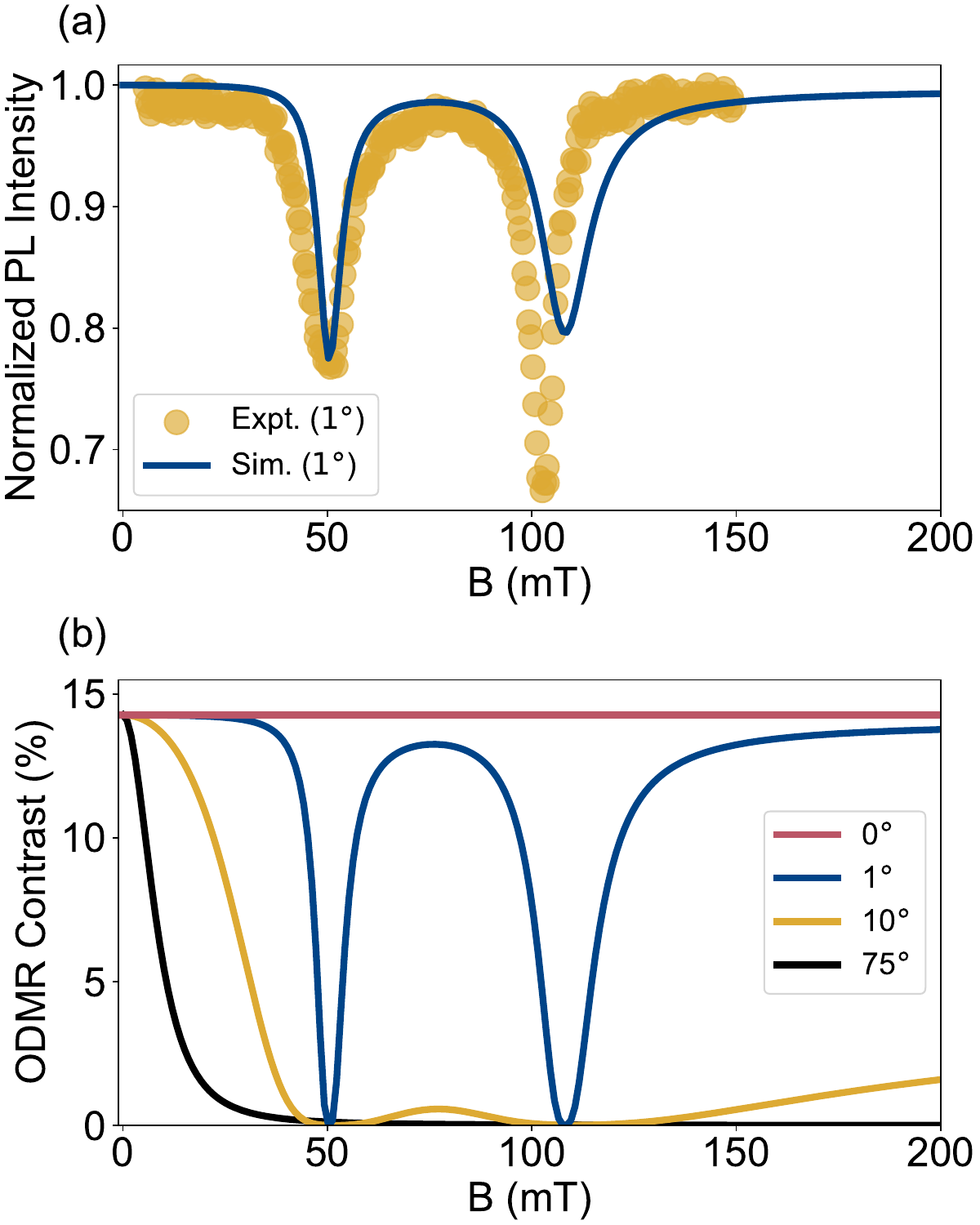}
    \caption{(a) Normalized PL intensity simulated using first-principles rates with spin-flip TDDFT data~\cite{jin2022vibrationally} against the applied magnetic field $\mathbf{B}$, with a direction of  $1\si{\degree}$ relative to the NV axis. The experiment for comparison is from Ref.~\cite{epstein2005anisotropic}. (b) ODMR contrast as a function of magnetic field with an angle relative to the NV axis [111], which is the spin quantization axis at $B=0$. The strong magnetic field dependency is originated from spin mixing. In both the normalized PL intensity and ODMR contrast, the sharp reduction at $B={\sim}50$ and ${\sim}100$ mT corresponds to level-anticrossing (LAC) due to ZFS and Zeeman effect in the excited state $\tes$ and ground state $\gs$, respectively.  At $\theta_B=0\si{\degree}$, the ODMR contrast is optimized with no spin mixing. }
    \label{fig:level-anticrossing}
\end{figure}

Finally, we study the ODMR contrast dependence on the magnetic field direction. How spin sublevels mix depends on the magnetic field direction with respect to the NV axis, which is quantified by polar angle $\theta_B$.
In Figs.~\ref{fig:level-anticrossing}(a) and \ref{fig:level-anticrossing}(b), we show both the angle dependency and magnetic-field dependency of normalized PL intensity and ODMR contrast. The simulated normalized PL intensity at $\theta_B = 1\si{\degree}$ exhibits similar sharp reduction character as the experiment~\cite{epstein2005anisotropic}. This reduction shares the same origin as the ODMR, as elaborated below. When the magnetic field is perfectly aligned with the NV axis, the ODMR contrast is maximal since there is no spin mixing. When magnetic field is slightly misaligned to the NV axis, we can see two positions of sharp reductions of the ODMR contrast at $B={\sim}50$ mT and ${\sim}100$ mT, which are related to the ZFS in $\tes$ and the one in $\gs$, respectively.
This is resulted from the excited state level-anticrossing (ESLAC) and the ground-state level-anticrossing (GSLAC). As can be seen in SM Fig.~\textcolor{red}{S21}~\cite{supplementary_material} that shows the extend to which the spin sublevels mix, there is strong spin mixing of $\ket{0}$ to $\ket{-1}$ at ESLAC and GSLAC. When the magnetic field is more misaligned with the NV axis, i.e. $10\si{\degree}<\theta_B<90\si{\degree}$, the ODMR contrast becomes highly sensitive to the magnetic field, nearly vanishing after the ESLAC. In addition, we find that the GSLAC and ESLAC of the NV center gradually disappear with increasing magnetic field after $B>25$ mT. The vanishing GSLAC and ESLAC can also be reflected by the ODMR frequency plots in  SM Fig.~\textcolor{red}{S22}~\cite{supplementary_material}. In general, this result suggests our theory can reliably predict the ODMR dependence on magnetic field direction, which is useful for setting up experiments. 
Considering the complexity of the computation process, it is beneficial to outline the procedures for ODMR simulation from first principles. Fig.~\ref{fig:workflow} illustrates the overall workflow.

\subsection{Optimization for ODMR Contrast}
The optimization of ODMR contrast is a critical step experimentally, as we show next can be realized through tuning optical saturation parameter $\beta$ (related to excitation efficiency $\beta k_R$) and tuning Rabi frequency $k_\text{MW}$~\cite{dreau2011avoiding}. Such parameters are cumbersome to search for experimentally but our theory can be much more efficient.  
%

Considering that the $k_{\perp/z}(\vibsgs\rightarrow\vibgs)$ rates largely vary across different experiments~\cite{robledo2011spin,tetienne2012magnetic}, and that the optimization behavior of the ODMR contrast is sensitive to these variations, it is essential to understand the relationship between the ISC rate and the optimization of ODMR. For clarity, we use the scale factor $k/k_0$ to represent the variations of $k_{\perp/z}(\vibsgs\rightarrow\vibgs)$, and $k_0$ denotes the $k_{\perp/z}(\vibsgs\rightarrow\vibgs)$ rates from our calculation in Table~\ref{tab:e-ph_ISC}.

\begin{figure}[h!]
    \centering
    \includegraphics[width=0.45\textwidth]{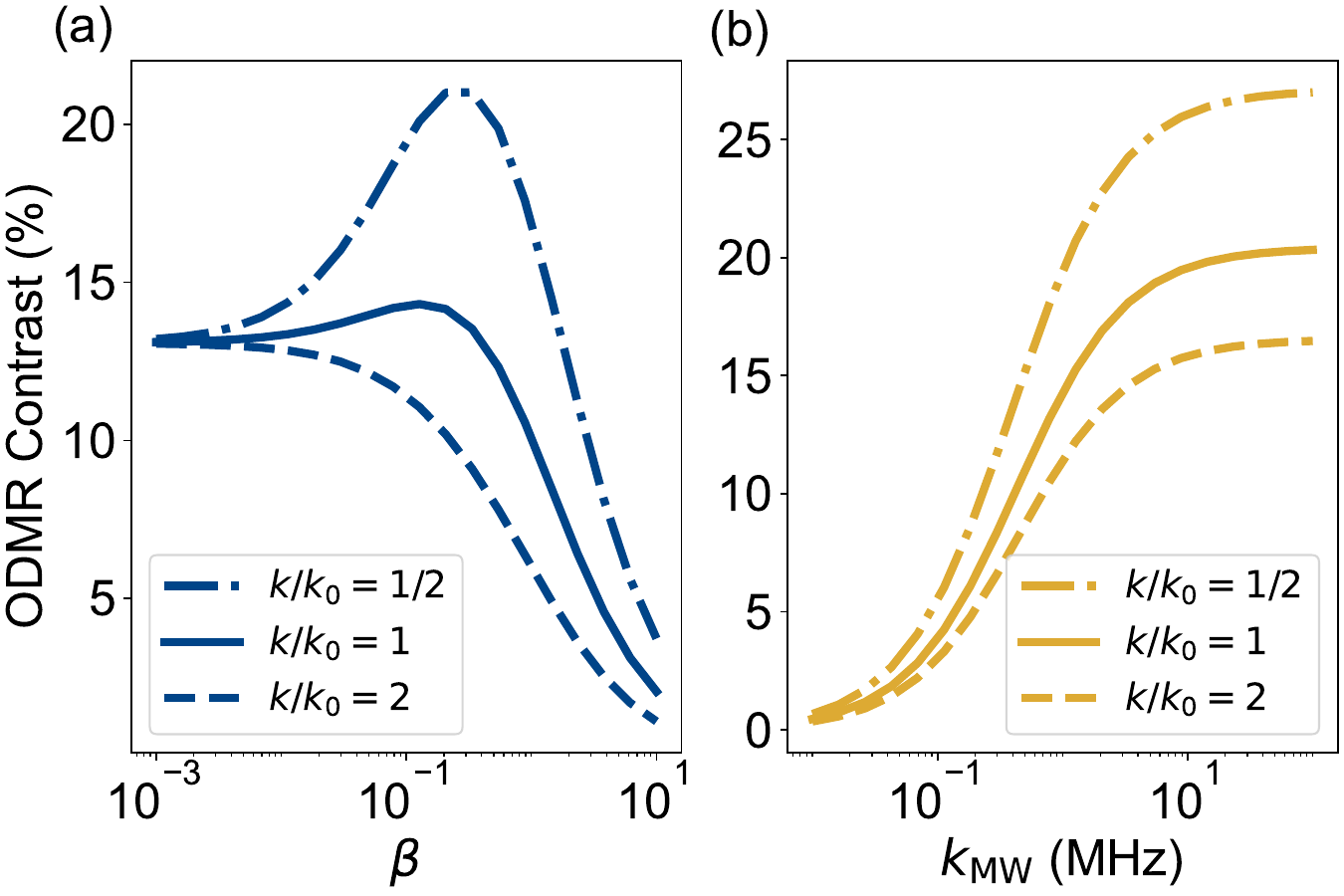}
    \caption{ODMR contrast can be optimized by tuning 
    (a) optical saturation parameter $\beta$ and 
    (b) Rabi frequency $k_\text{MW}$. The ODMR is simulated at $B=0$ and $\theta_B=0$. The  corrected rates by using the spin-flip TDDFT data are used for the ODMR simulation to show the optimization of ODMR contrast. $k_0$ denotes the $k_{\perp/z}(\vibsgs\rightarrow\vibgs)$ rates from our calculation. With maintaining the ratio of $k_{\perp}(\vibsgs\rightarrow\vibgs)/k_{z}(\vibsgs\rightarrow\vibgs)$, the scaled version, $k$, is used to illustrate how the optimization behavior of ODMR contrast changes with the variation in the $k_{\perp/z}(\vibsgs\rightarrow\vibgs)$ rates.
    }
    \label{fig:optimize_odmr}
\end{figure}

Fig.~\ref{fig:optimize_odmr}(a) shows that the ODMR is optimized at $\beta<0.01$ when the $k_{\perp/z}(\vibsgs\rightarrow\vibgs)$ rates are overestimated compared to the experimental observation. If the $k_{\perp/z}(\vibsgs\rightarrow\vibgs)$ rates are close to or below experimental values, the ODMR is optimized when the optical excitation power is ${\sim0.1}$ of the saturation. Despite variations in the $k_{\perp/z}(\vibsgs\rightarrow\vibgs)$ rates, the part in $\beta>0.1$ is consistent with the experiment~\cite{dreau2011avoiding}. However, the high sensitivity of the ODMR optimization to $\beta$ underscores the importance of obtaining accurate rates for the correct prediction. 

Fig.~\ref{fig:optimize_odmr}(b) indicates that the ODMR contrast can reach its maximum at $k_\text{MW}\approx 5$ MHz. The variations in the $k_{\perp/z}(\vibsgs\rightarrow\vibgs)$ rates primarily affect the magnitude of the ODMR contrast with respect to $k_\text{MW}$. When the populations of the spin sublevels are driven by the Rabi oscillation, the nonaxial ISC $\vibtes\rightarrow\vibfses$ becomes the preferential path for the relaxation from $\vibtes$. Thus, the upper bound of the ODMR contrast in this case is limited by the contrast between $k_\perp(\vibtes\rightarrow\vibfses)$ and $k_z(\vibtes\rightarrow\vibfses)$.

\section{Conclusion}
\begin{figure}[h!]
\centering
\begin{tikzpicture}[node distance=2cm]
\tikzstyle{startstop} = [rectangle, rounded corners, minimum width=3cm, minimum height=1cm,text centered, draw=black, fill=myblue!30]
\tikzstyle{process1} = [rectangle, rounded corners, minimum width=2cm, minimum height=1cm, text centered, draw=black, fill=myorange!30]
\tikzstyle{process2} = [rectangle, rounded corners, minimum width=2cm, minimum height=1cm, text centered, draw=black, fill=myred!30]
\tikzstyle{arrow} = [thick,->,>=stealth]
\node (start) [startstop] {\makecell{Electronic Structure and Optimized Geometry \\(e.g. DFT, spin-flip TDDFT)}};
\node (process1a) [process1, below left of=start, xshift=-1.5cm] {\makecell{VEE+$\mu_{e-h}$\\ (e.g. DFT, $\gwbse$)}};
\node (process1b) [process1, below of=start, yshift=-0.8cm, xshift=-1.5cm] {\makecell{SOC\\ (e.g. CASSCF)}};
\node (process1c) [process1, below of=start, yshift=-0.8cm, xshift=1.5cm] {e-ph coupling};
\node (process1d) [process1, below right of=start, xshift=1.5cm] {ZFS};
\node (process2a) [process2, below of=process1a, yshift=-2.7cm] {$k_\text{R}$};
\node (process2b) [process2, below of=process1b] {$k_\text{ISC}$};
\node (process2c) [process2, below of=process1c] {$k_\text{IC}$};
\node (process2d) [process2, below of=process1d, yshift=-2.7cm] {$\alpha(\mathbf{B})$};
\node (stop) [startstop, below of=start, yshift=-5cm] {ODMR Simulation};
\draw [arrow] (start) -- (process1a);
\draw [arrow] (start) -- (process1b);
\draw [arrow] (start) -- (process1c);
\draw [arrow] (start) -- (process1d);
\draw [arrow] (process1a) -- (process2a);
\draw [arrow] (process1b) -- (process2b);
\draw [arrow] (process1c) -- (process2b);
\draw [arrow] (process1c) -- (process2c);
\draw [arrow] (process1d) -- node[anchor=west] {+ Zeeman} (process2d);
\draw [arrow] (process2a) -- (stop);
\draw [arrow] (process2b) -- (stop);
\draw [arrow] (process2c) -- (stop);
\draw [arrow] (process2d) -- (stop);
\end{tikzpicture}
\caption{Workflow of ODMR simulation from first-principles. VEE stands for vertical excitation energy. $\mu_{e-h}$ represents optical dipole moment. e-ph coupling represents electron-phonon coupling. $k$ represents the rates which can be obtained using the Fermi's golden rule. $\alpha(\mathbf{B})$ represents the mixing coefficient in Eq.~(\ref{eq:B_dependent_rate}).}
\label{fig:workflow}
\end{figure}
In conclusion, we have developed general first-principles computational platform for spin-dependent PL contrast and cw-ODMR, for triplet and singlet spin defects in solids. We solved the kinetic master equation to obtain steady-state excited-state occupations. These occupations are determined by excited-state kinetic rates, which we implemented fully from first-principles. We then validated our implementation by comparing with the experimental data on NV center in diamond. We show our first-principles computational platform for ODMR is fully general and accurate, as long as we have reliable electronic structure as inputs (including state energies, wavefunction, and defect geometry).

The ODMR simulation from first principles requires  calculations of spin sublevels due to ZFS, and the rates of all possible transitions i.e. radiative recombination, ISC (spin-flip nonradiative recombination) and IC (spin-conserving nonradiative recombination). In this work we examined the theory and implementation of ISC in great detail, emphasizing accurate description of both spin-orbit coupling and electron-phonon coupling.  
In particular, by using NV center in diamond as prototypical example, we identify the coupling of $\sgs$ and $\sses$ due to the configuration interaction 
from CASSCF calculation and group theory, and show the corresponding effect on SOC matrix elements. We complete the derivation and calculation for the effective SOC matrix elements of the NV center considering the pseudo JT effect. 
As a result, we unequivocally clarify the experimental observation of the axial ISC $\vibtes\rightarrow\vibfses$, which was thought to be forbidden in the first order by symmetry. Importantly, we show the dynamical JT effect in the degenerate states $\tes$ and $\sgs$ significantly enhances the nonradiative recombination by reducing the potential energy barrier. The results  showcase the rich physics underlying the entire excitation and relaxation cycle. 

Finally, we simulate cw-ODMR of the NV center from first-principles calculations. Through our calculations, we attribute the dip in ODMR contrast at $B=0$ to the non-zero rhombic ZFS parameter $E$, indicating symmetry breaking of the NV center in the experiment.
We provide the optimization strategies for ODMR contrast with respect to magnetic field, 
optical saturation parameter and Rabi frequency. These are found to be informative especially for the range of conditions difficult to reach experimentally. 
This shows the critical role of our computational technique can play in guiding experiments, which also enables deeper understanding to the spin polarization of spin defects.

The study emphasizes the importance of accounting for the multi-reference character of the electronic states, SOC and electron-phonon coupling for spin defects. It unveils the challenge that there is a need of developing advanced first-principles theory for accurate prediction of SOC and electron-phonon coupling in solid-state spin defects. 
On the other hand, the 
atomic structure of many spin defects has not been determined yet, such as ST1 defect in diamond and quantum defects in two-dimensional wide-band gap semiconductors. Our developed computational platform can be essential for identifying existing spin defects, and potentially applied to the design of new solid-state spin defect important for nanophotonics and quantum information science.


\begin{acknowledgments}
We acknowledge Susumu Takahashi and Yu Jin for very helpful discussions on ODMR physics and spin-flip TDDFT. Ping, Li, and Zhang acknowledge the support by the National Science Foundation under grant no. DMR-2143233.
Varganov, V. Dergachev, and I. Dergachev acknowledge the support by the U.S. Department of Energy, Office of Science, Office of Basic Energy Sciences Established Program to Stimulate Competitive Research under Award Number DE-SC0022178.
This research used resources of the Scientific Data and Computing Center, a component of the Computational Science Initiative, at Brookhaven National Laboratory under Contract No. DE-SC0012704,
the Lux Supercomputer at UC Santa Cruz, funded by NSF MRI Grant No. AST 1828315,
the National Energy Research Scientific Computing Center (NERSC), a U.S. Department of Energy Office of Science User Facility operated under Contract No. DE-AC02-05CH11231,
and the Extreme Science and Engineering Discovery Environment (XSEDE), which is supported by National Science Foundation Grant No. ACI-1548562 \cite{xsede}.
\end{acknowledgments}

\bibliography{ref}
\end{document}